\documentclass[journal,twoside,web]{ieeecolor}
\usepackage{generic}
\usepackage{cite}
\usepackage{amsmath,amssymb,amsfonts}
\usepackage{algorithmic}
\usepackage{graphicx}
\usepackage{textcomp}
\usepackage{url}
\usepackage{hyperref}
\usepackage{lineno}
\def\BibTeX{{\rm B\kern-.05em{\sc i\kern-.025em b}\kern-.08em
    T\kern-.1667em\lower.7ex\hbox{E}\kern-.125emX}}
\begin{document}
\title{Circuit-aware Device Modeling of Energy-efficient Monolayer WS$_2$ Trench-FinFETs}

\author{Tarun Agarwal, Youseung Lee, and Mathieu Luisier
\thanks{T. Agarwal is currently with Integrated Systems Laboratory, ETH Zurich, Switzerland and will soon be joining the Department
of Electrical Engineering, IIT Gandhinagar, Gandhinagar, India (e-mail:tagarwal@iitgn.ac.in).}
\thanks{Y.~Lee and M.~Luisier are with Integrated Systems Laboratory, ETH Zurich, Switzerland.}}

\maketitle

\begin{abstract}
The continuous scaling of semiconductor technology has pushed the footprint of logic devices below 50 nm. Currently, logic standard cells with one single fin are being investigated to increase the integration density, although such options could severely limit the performance of individual devices. In this letter, we present a novel Trench (T-) FinFET device, composed of a monolayer two-dimensional (2D) channel material. The device characteristics of a monolayer WS$_2$-based T-FinFET are studied by combining the first-principles calculations and quantum transport (QT) simulations. These results serve as inputs to a predictive analytical model. The latter allows to benchmark the T-FinFET with strained (s)-Si FinFETs in both quasi-ballistic and diffusive transport regimes. The circuit-level evaluation highlights that WS$_2$ T-FinFETs exhibit a competitive energy-delay performance compared to s-Si FinFET and WS$_2$ double-gate transistors, assuming the same mobility and contact resistivity at small footprints.
\end{abstract}

\begin{IEEEkeywords}
2D materials, monolayer WS$_2$ FinFET, multi-scale modeling, benchmarking, strained-Si FinFETs.
\end{IEEEkeywords}

\section{Introduction}
\label{sec:introduction}
\IEEEPARstart{T}{he} technology scaling has necessitated the downsizing of contacted gate pitch (CGP) and minimum footprint (W$_{foot}$) of contemporary FinFETs to achieve Power-Performance-Area (PPA) requirement of sub-10 nm technology nodes \cite{OneFinFET_option}. As the next technology option, one-fin device might be considered for their potential to reduce the footprint. In such configurations, only few process parameters can control the performance of FinFETs, for example: the fin height (H$_{fin}$). Alternatively, a different channel material with better transport properties than silicon could be used \cite{2DFinFET_exp}.

Recently, two-dimensional (2D) materials have emerged as promising candidates to build the next-generation high-performance transistors thanks to their attractive electrical and mechanical properties \cite{2D_microprocessor}. Among them, monolayer WS$_2$, a so-called transition-metal-dichalcogenide (TMD), has attracted considerable attention from the semiconductor community. Indeed, WS$_2$-based transistors promise high ON-state currents owing to the low transport effective mass, high intrinsic carrier mobility \cite{WS2_Mobility,Youseung_IEDM}, and excellent electrostatics associated with the 2D channel. Hence, monolayer WS$_2$ shows great prospects for future sub-10 nm gate length transistors \cite{Tarun_SciRep}.

\begin{figure}[!t]
\centerline{\includegraphics[width=\columnwidth]{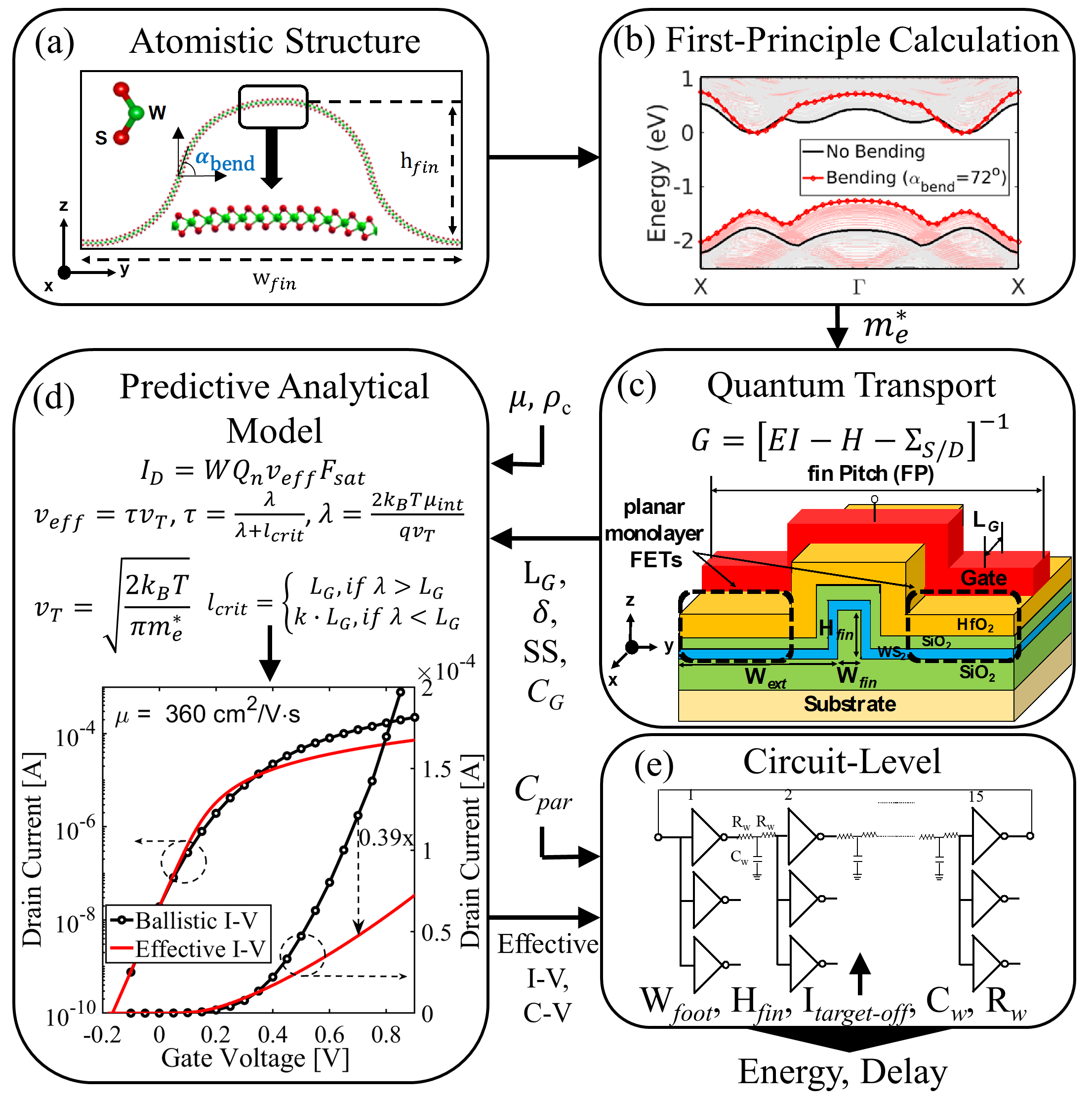}}
\caption{\textbf{Methodology} a) Creation of a bent atomic monolayer WS$_2$ structure mimicking its deposition on a SiO$_2$ fin, b) Electronic bandstructure calculation of the atomic cell with and without bending using first-principles calculations, c) Simulation of the monolayer WS$_2$-based T-FinFET over one fin pitch with a ballistic QT solver, d) Extraction of effective gate capacitance ($C_G$), sub-threshold slope (SS) and drain-induced barrier lowering ($\delta$), from QT calculations and insertion of them into a predictive analytical model \cite{Tarun_IEDM} to obtain the ``I-V" characteristics of the T-FinFET, e) Ring-oscillator simulation with the analytical model of \cite{MVSmodel} and derivation of relevant circuit-level metrics such as delay and energy-consumption.}
\label{fig:1}
\end{figure}

In this letter, we investigate the potential of an alternative 2D material device, the WS$_2$ Trench (T-) FinFET, at a device footprint of 24 nm. 
Its electrical behavior is evaluated through first-principles and quantum transport (QT) calculations combined with circuit-level simulations, as explained in Section~\ref{Methodology}. The electronic properties of the fin-shaped monolayer WS$_2$ are first obtained from density functional theory \cite{VASP} and then passed to a ballistic QT solver \cite{QTSolver} to calculate the current vs. voltage and capacitance vs. voltage characteristics. In Section~\ref{Results}, we benchmark the energy efficiency of the WS$_2$ T-FinFET with the WS$_2$ double-gate FET and strained (s)-Si FinFET studied in our previous work \cite{Tarun_TED}. The WS$_2$ T-FinFET shows lower energy consumption and delay than s-Si FinFET for the same channel mobility.


\section{Methodology}
\label{Methodology}
\par Our circuit-aware device modeling methodology is outlined in Fig.~\ref{fig:1}. First, the monolayer WS$_2$ is bent with an angle ($\alpha_{bend}$) to generate the required T-Fin structure, which resembles the conventional Fin shape (Fig.~\ref{fig:1}(a)). The electronic structure of the bent monolayers is then computed from the first-principles VASP simulator \cite{VASP} and the generalized gradient approximation (GGA) of Perdew, Burke, and Ernzerhof (PBE) \cite{PBE}. 
Table~\ref{effective mass table} summarizes the width and height of the considered fin-shaped WS$_2$ monolayers (w$_{fin}$ and h$_{fin}$), and the corresponding electron/hole effective masses (m$^*_e$/m$^*_h$) and bandgap (E$_G$) for different $\alpha_{bend}$ values. It is worth noting that the band curvature or m$^*_e$ at the conduction band minimum (CBM) remains unchanged, regardless of the bending angle, while the location of the valence band maximum (VBM) changes from K-valley to $\Gamma$-valley resulting in a change in E$_G$, as shown in Fig.~\ref{fig:1}(b). This finding allows us to safely resort to the effective mass approximation to perform quantum transport simulations within the non-equilibrium Green's function (NEGF) formalism, thus making the investigation of realistic device structures possible.

\begin{table}[!t]
\renewcommand{\arraystretch}{1.3}
\caption{Material parameters extracted from first-principles calculations}
\label{effective mass table}
\centering
\begin{tabular}{|c|c|c|c|c|c|}
\hline
$\alpha_{bend}$ & w$_{fin}$ & h$_{fin}$ & m$^*_e$ (Valley) & m$^*_h$ (Valley) & E$_G$ \\
($^{\circ}$) & (${\AA}$) & (${\AA}$) & (* m$_0$) & (* m$_0$) & (eV) \\
\hline
0 & 285 & -- & 0.33 (K) & 0.5 (K) & 1.75 \\
\hline
48 & 261 & 62 & 0.32 (K) & 2.0 ($\Gamma$) & 1.22  \\
\hline
60 & 241 & 74 & 0.32 (K) & 2.0 ($\Gamma$) & 1.27  \\
\hline
72 & 219 & 84 & 0.32 (K) & 2.0 ($\Gamma$) & 1.26  \\
\hline
72 & 424 & 160 & 0.32 (K) & 2.0 ($\Gamma$) & 1.09  \\
\hline
\end{tabular}
\end{table}

\par Fig.~\ref{fig:1}(c) shows the schematic of the WS$_2$ T-FinFET with a fin Pitch (FP) = 24 nm and W$_{fin}$ = 5 nm. Here, the device simulation domain in the width direction is set to one FP. The quasi-ballistic/diffusive device characteristics are then calculated by combining the ballistic QT simulations and the predictive analytical model, shown in Fig.~\ref{fig:1}(d), to account for the scattering in the WS$_2$ channel. In this approach, the key device parameters are first extracted from ballistic QT, e.g. the effective gate capacitance ($C_G$), sub-threshold slope (SS), and drain-induced barrier lowering ($\delta$). The extracted parameters are  then injected into the predictive analytical model to calculate the effective drain currents. Fig.~\ref{fig:1}(d) also compares the ballistic and quasi-ballistic transfer characteristics of monolayer WS$_2$ T-FinFET that are obtained from QT simulations and from the predictive analytical model, respectively. The ratio of effective ON current (I$_{\rm{ONeff}}$) and ballistic ON current (I$_{\rm{ONball}}$) is indicated to be 0.39 at V$_{DD}$= 0.7 V. Interestingly, as indicated in Fig.~\ref{fig:1}(c), WS$_2$ T-FinFET comprises of two additional planar monolayer WS$_2$ FETs in the trench. It then shows more effective device width ($W_{dev}$ = 2*(W$_{ext}$+H$_{fin}$)+W$_{fin}$) within one FP than s-Si FinFETs (2*H$_{fin}$+W$_{fin}$). For a FP of 24 nm, the maximum width of these planar FETs is 6.8 nm, taking into account the 0.7 nm thickness of monolayer WS$_2$, 0.5 nm of the SiO$_2$ interfacial layer, and 1.5 nm of the HfO$_2$ dielectric layer. 

Finally, to assess the circuit-level metrics, we perform the ring-oscillator (RO) simulations including the device and interconnect parasitics, using Cadence Spectre Simulation Platform \cite{Cadence}. The parasitic capacitances ($C_{par}$) depend on the FET architectures and are calculated using the analytical expressions proposed in \cite{JLacord_paper,Tarun_TED2}. The wire capacitance (C$_w$ = 0.27 fF/$\mu$m) and resistances (R$_w$ = 317 $\Omega$/$\mu$m) are included between each RO stage in the circuit-level simulation, with a wire length of 50 CGP, as shown in Fig.~\ref{fig:1}(e). The contact resistivity ($\rho_C$) is chosen to be 10$^{-8}$ $\Omega$.cm$^2$, which corresponds to a contact resistance of 100 $\Omega.\mu$m for a contact length of 10 nm. We also assume balanced nFET and pFET performance in the RO simulation. Hence, the circuit evaluation primarily benchmarks the n-channel devices. To show the energy-efficiency of different device options, the energy-delay product (EDP) is chosen as the central figure-of-merit in this letter.

\section{Results and Discussion}
\label{Results}
\begin{figure}[htp]
	\centering
        \includegraphics[scale=0.40]{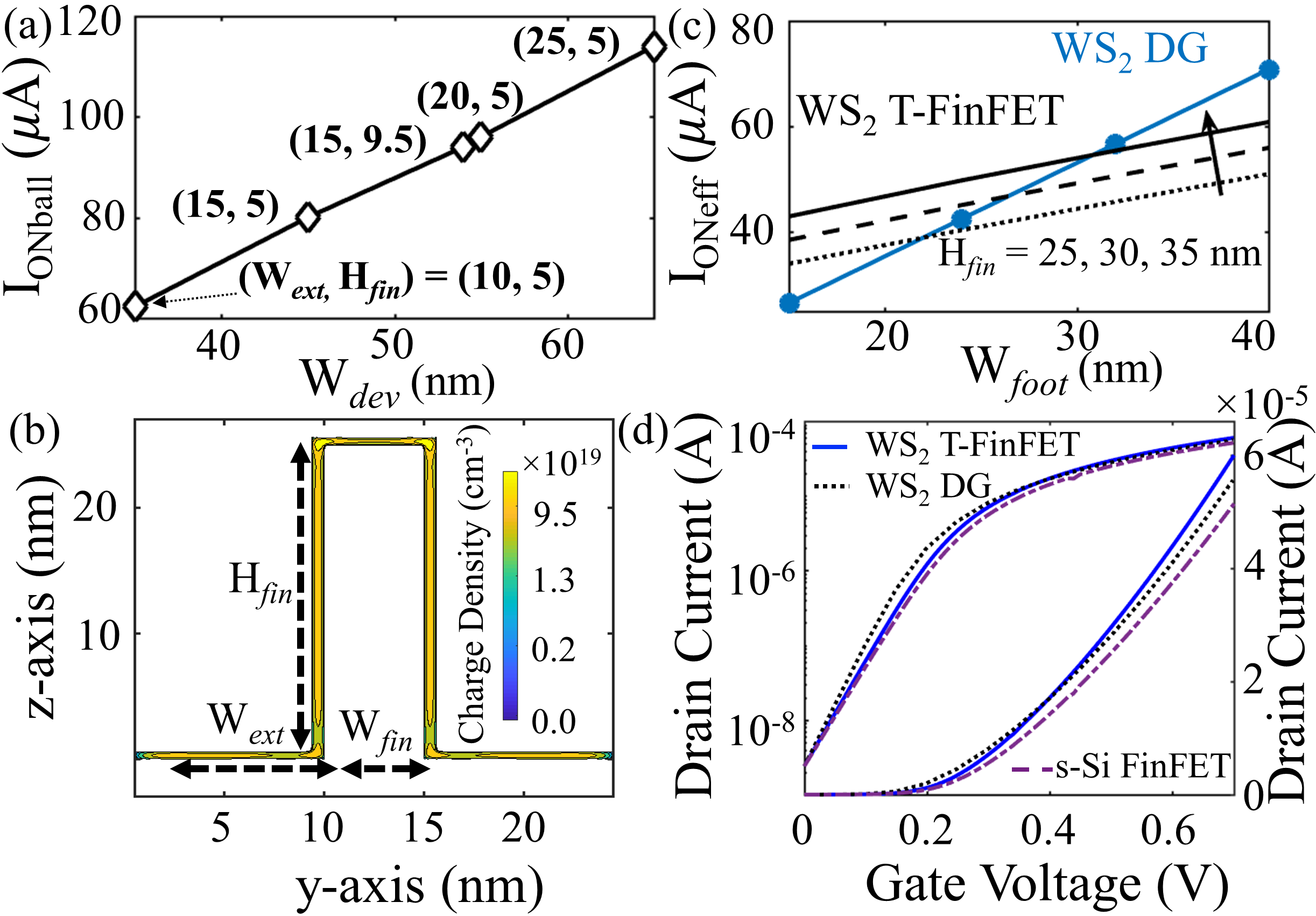}
		\caption[]{\textbf{Device characteristics}, a) Ballistic ON current with effective device width (W$_{dev}$) for different H$_{fin}$ and W$_{ext}$, b) Charge density in the WS$_2$ channel distributed across the T-FinFET width (y) and height (z) at V$_{GS}$= 0.61 V and V$_{DS}$= 10 mV, c) Comparison of the effective ON current of a WS$_2$-based T-FinFET and double-gate (DG) FET as a function of the device footprint (W$_{foot}$). Here, I$_{OFF}$ is set to 100 nA/$\mu$m (2.4 nA for FP=24 nm), d) Comparison of the I$_D$-V$_G$ characteristics of a WS$_2$-based T-FinFET and DGFET as well as s-Si FinFET at V$_{DD}$= 0.61 V. The source and drain extension lengths are set to 15 nm for all cases.}
	\label{fig:2}
\end{figure}

Fig.~\ref{fig:1}(d) shows the transfer characteristics of WS$_2$ T-FinFET for a fixed FP = 24 nm and H$_{fin}$ = 15 nm, with an intrinsic phonon-limited electron mobility $\mu$ = 360 cm$^2$/(V$\cdot$s) \cite{Youseung_IEDM}. The effective device width can be controlled by changing either the fin height (H$_{fin}$) or the extensions of the channel width (W$_{ext}$). Only a change in W$_{ext}$ affects the FinFET footprint (W$_{foot}$ = FP) while H$_{fin}$ does not. Fig.~\ref{fig:2}(a) exhibits that the I$_{\rm{ONball}}$ obtained from QT framework scales linearly with W$_{dev}$, i.e. with both H$_{fin}$ and W$_{ext}$. This is due to the rather uniform distribution of the charge density across the fin-shaped WS$_2$ channel, as highlighted in Fig.~\ref{fig:2}(b).

\par Fig.~\ref{fig:2}(c) compares the I$_{\rm{ONeff}}$ of WS$_2$-based T-FinFETs with different fin heights to that of double-gate (DG) FET for device footprints (W$_{foot}$) ranging between 15 and 40 nm. It appears that the DG FETs are only beneficial at footprints larger than 30 nm. By increasing H$_{fin}$, the T-FinFET can always outperform the DG FET, demonstrating the advantage of the proposed architecture. Next, we compare the transfer characteristics of the WS$_2$-based T-FinFET and DG FET to those of a conventional s-Si FinFET in Fig.~\ref{fig:2}(d), using $C_G$, SS and $\delta$ values listed in Table~\ref{device properties table}. Thanks to their atomic layer thickness, the T-FinFET and DG-FET show a better gate controllability than the s-Si FinFET. However, the performance of these device architectures need to be compared at the circuit-level by combining the device characteristics with parasitic capacitances and resistances.

\begin{table}[!t]
\renewcommand{\arraystretch}{1.3}
\caption{Key input and extracted device parameters}
\label{device properties table}
\centering
\begin{tabular}{|c|c|c|c|c|c|c|}
\hline
Device & L$_G$ & $\mu$ & $C_G$ & SSsat & $\delta$ \\
 & (nm) & (cm$^2$/V.s) & ($\mu$F/cm$^2$) & (mV/dec) & (V/V) \\
\hline
WS$_2$ DG & 12  & 360 & 6.4 & 60.76 & 0.05  \\
\hline
WS$_2$ T-FinFET & 12 & 360 & 2.9 & 70.9 & 0.075 \\
\hline
s-Si FinFET & 12 & 680 & 2.5 & 75.28 & 0.1 \\
\hline
\end{tabular}
\end{table}

\par
\begin{figure}[htp]
	\centering
        \includegraphics[scale=0.55]{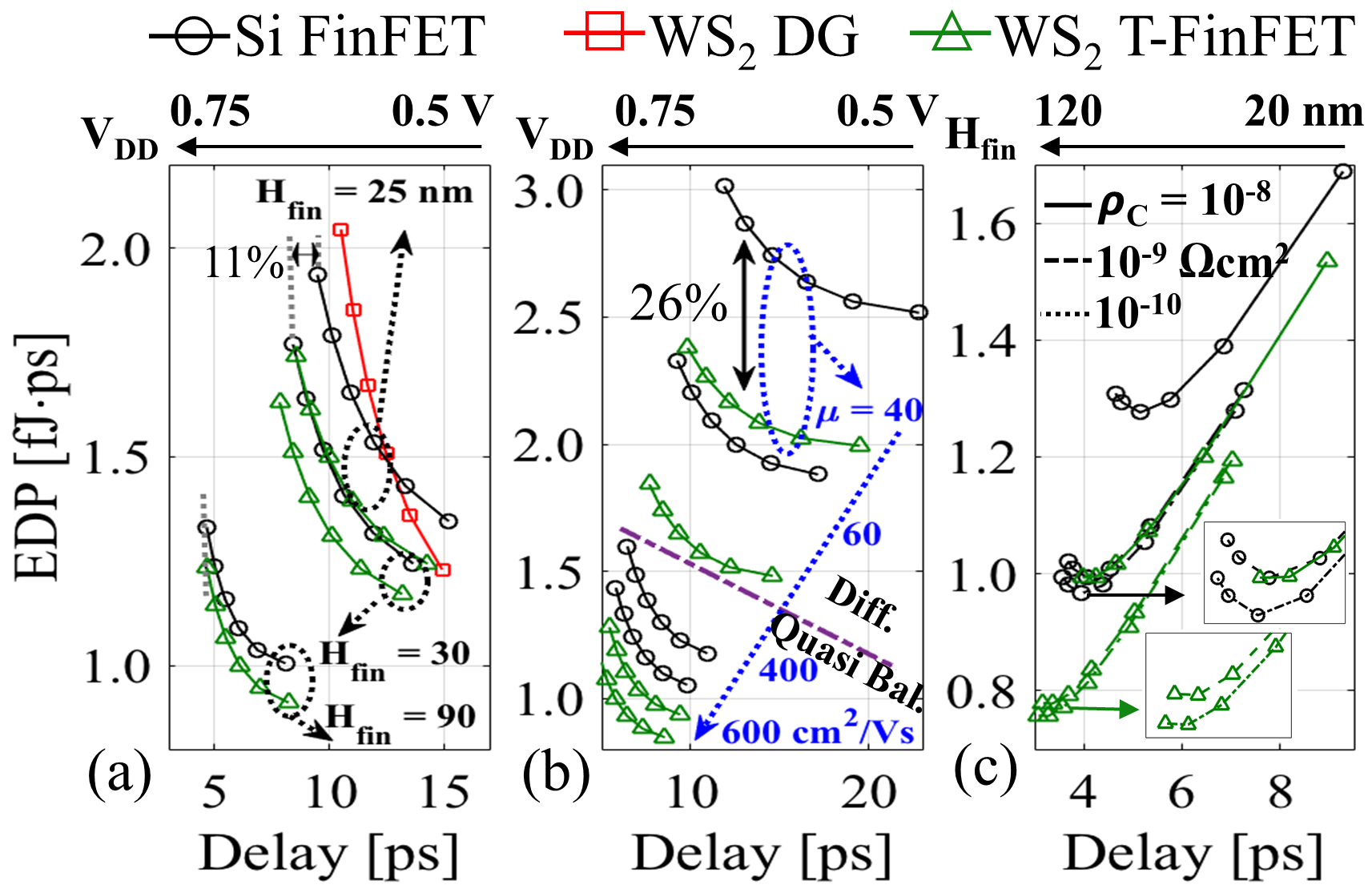}
		\caption[]{\textbf{Circuit-level benchmarking} of WS$_2$ T-FinFET, DG FET, and s-Si FinFET for CGP= 42 nm, a spacer length of 10 nm and permittivity of 3.9. The Energy-Delay Product (EDP) vs. delay is shown as a function of (a) V$_{DD}$ for different H$_{fin}$ at $\mu_{WS_2(s-Si)}$= 360 (680) cm$^2$/Vs and $\rho_C$= 10$^{-8}$ $\Omega$.cm$^2$, (b) V$_{DD}$ for different mobility values at H$_{fin}$= 60 nm and $\rho_C$= 10$^{-8}$ $\Omega$.cm$^2$, and (c) H$_{fin}$ for different contact resistivity values at $\mu$= 600 cm$^2$/Vs and V$_{DD}$= 0.7 V.}
	\label{fig:3}
\end{figure}

 Fig.~\ref{fig:3} shows the ``EDP vs delay" between WS$_2$ T-FinFET, DG FET, and s-Si FinFET for different H$_{fin}$, supply voltage (V$_{DD}$), and contact resistivity ($\rho_C$). In this study, the bottom-left corner of each plot represents the sweet spot corresponding to a more energy-efficient and fast-switching device operation. In Fig.~\ref{fig:3}(a), we see that the WS$_2$ DG FET provides a higher energy-efficiency (i.e. lower EDP) than both the WS$_2$ T-FinFET and s-Si FinFET of H$_{fin}$= 25 nm for V$_{DD}$ below 0.55 V. However, by increasing H$_{fin}$ without enlarging the footprint, both the WS$_2$ T-FinFET and s-Si FinFET always outperform the WS$_2$ DG FET, regardless of V$_{DD}$. From Fig.~\ref{fig:3}(a), it can also be observed that s-Si FinFET shows almost the same delay as the WS$_2$ T-FinFET at H$_{fin}$= 90 nm while a 11 $\%$ higher delay at H$_{fin}$= 25 nm. This is due to the fact that the influence of the lateral WS$_2$ extensions of T-FinFETs (see Fig.~\ref{fig:1}(c)) vanishes at large H$_{fin}$.

\par Going one step further, to understand the role of the channel mobility ($\mu$), we compare the WS$_2$ T-FinFET and s-Si FinFET in the quasi-ballistic (high $\mu$) and diffusive (low $\mu$) regimes, as shown in Fig.~\ref{fig:3}(b). The WS$_2$ T-FinFET appears as a better device option than the s-Si FinFET at constant channel mobility and contact resistance. 
At small mobility values, e.g. 40 cm$^2$/Vs, which is the result of charged impurity scattering in WS$_2$ with an impurity concentration N$_{imp}$=1.35x10$^{12}$ cm$^{-2}$ \cite{Youseung_IEDM}, the EDP of the WS$_2$ T-FinFET is 26 $\%$ lower than that of the s-Si FinFET for the same delay. The advantage of T-FinFET is about the same (28 $\%$ lower EDP) at higher mobility values, i.e. 600 cm$^2$/Vs.
\par The effect of $\rho_C$ on the EDP is highlighted in Fig.~\ref{fig:3}(c) at different H$_{fin}$. The insets show a larger EDP increase at high H$_{fin}$ for the s-Si FinFET than the WS$_2$ T-FinFET due to its higher total capacitance. Also, $\rho_C$ values below 1x10$^{-9}$ $\Omega$.cm$^2$ do not have any significant impact on the EDP, while a change in $\rho_C$ from 1x10$^{-8}$ to 1x10$^{-9}$ $\Omega$.cm$^2$ results in 18 and 23 $\%$ lower EDPs for the WS$_2$ T-FinFET and s-Si FinFET, respectively.

\section{Conclusion}
The electron/hole effective masses and bandgaps of fin-shaped monolayer WS$_2$ have been computed using the first-principles. It is found that the m$^*_e$ does not change with the bending angle. For the benchmarking, the energy-delay product of the WS$_2$ T-FinFET has been compared to the WS$_2$ DGFET as well as the s-Si FinFET through a circuit-aware device modeling framework. It has been shown that the WS$_2$ T-FinFET can outperform the s-Si FinFET with the same channel mobility and contact resistivity. For supply voltages above 0.5 V, the WS$_2$ T-FinFET promises better circuit-level performance than the WS$_2$ DGFET.

\end{document}